\documentclass[twoside,slac_one]{revtex4}
\usepackage{graphicx}
\usepackage{fancyhdr}
\usepackage{amsmath} % American Mathematics Society standards
\usepackage{bm}% bold math
\usepackage{amsxtra}
\usepackage{amssymb}
\usepackage{amsthm}
\usepackage{latexsym}
\usepackage{lscape}

\newcommand{\dm}{$\Delta m^{2}$}
\newcommand{\sn}{$\sin^2 2\theta$}
\newcommand{\eps}{$\epsilon$}

\begin{document}

\title{Search for Flavor Changing Non-standard Interactions with the MINOS Experiment}


\author{Zeynep Isvan, for the MINOS Collaboration}
\affiliation{Department of Physics and Astronomy, University of Pittsburgh, Pittsburgh, PA, USA}

\begin{abstract}

MINOS searches for neutrino oscillations using the disappearance of muon neutrinos between two detectors, over a baseline of 735 km.  We recently reported the most precise measurement of neutrino oscillations in the atmospheric sector and the first tagged measurement of antineutrino oscillations.  The neutrino mass splitting and mixing angle are measured to be $|\Delta m^{2}| = 2.32_{-0.08}^{+0.12} \times 10^{-3}\,eV^{2}$ and $\sin^{2}2\theta > 0.90$ (90\% C.L.) for an exposure of $7.25 \times 10^{20}$ protons-on-target (PoT).  Antineutrino oscillation parameters are measured as  $\Delta \overline{m}^{2}=(3.36^{+0.46}_{-0.40}\textrm{(stat.)}\pm0.06\textrm{(syst.)})\times 10^{-3}\,eV^{2}$ and $\sin^{2}(2\overline{\theta})=0.86^{+0.11}_{-0.12}\textrm{(stat.)}\pm0.01\textrm{(syst.)}$ with an exposure of $1.7 \times 10^{20}$ PoT in NuMI antineutrino running mode.  We use the apparent difference in neutrino and antineutrino oscillation parameters to constrain non-standard matter interactions which could occur during propagation through the Earth's crust to the Far Detector.
\end{abstract}

\maketitle

\thispagestyle{fancy}


\section{Introduction}
MINOS observes a difference between neutrino and antineutrino standard oscillation parameters \cite{ccprl2010, rhcprl2010}. The statistical probability that the neutrino and antineutrino data have the same oscillation parameters is 2\% \cite{rhcprl2010, Himmel:CombinedSig}. So-called non-standard matter interactions (NSI) of muon neutrinos and antineutrinos with the Earth's crust during their propagation can be a possible explanation of this difference, since neutrino (matter) and antineutrino (antimatter) survival probabilities are altered in opposite directions by this effect. MINOS is particularly well suited to perform this measurement since its detectors are magnetized and neutrinos and antineutrinos can be distinguished on an event-by-event basis.  

This document details the non-standard interactions (NSI) analysis which uses a $7.09 \times 10^{20}$ POT neutrino sample and a $1.71 \times 10^{20}$ POT antineutrino sample. In this analysis neutrino and antineutrino vacuum oscillation parameters, \dm\ and \sn\ are assumed identical and the entire difference between the observed rate of disappearance is attributed to non-standard interactions. 

\subsection{Neutrino Oscillations}
Neutrino oscillations occur because neutrino flavor eigenstates are not identical to the mass eigenstates but instead are related by the Pontecorvo-Maki-Nakagawa-Sakata (PMNS) mixing matrix, $U$. The probability that a neutrino of energy $E$ created with flavor $\alpha$ will be detected in a flavor state $\beta$ after traveling a distance $L$ is given by

\begin{equation}
P\left(\nu_{\alpha}\rightarrow\nu_{\beta}\right)=
\left| 
\sum_{j=1}^{3}U^{*}_{\alpha j}\exp\left( -\frac{im^{2}_{j}L}{2E}  \right)U_{\beta j}
\right|^{2},
\end{equation}
where $m_{j}$ is the mass of the $j$th mass eigenstate. The final expression for oscillation probability depends not on the absolute masses of the neutrinos but on the squared difference between masses, $\Delta m^{2}_{32}$ and $\Delta m^{2}_{21}$. Because of the large difference between the magnitudues of these two mass squared splittings, experiments such as MINOS are sensitive to mixing between two of the three flavors. In the two-flavor approximation the survival probability for a muon neutrino is given by:

\begin{equation}
P(\nu_{\mu}\rightarrow\nu_{\mu}) = 1 - \sin^{2}2(\theta_{23})
\sin^{2}\left( 1.27\frac{\Delta m^{2}_{32}L}{E}  \right)
\end{equation}

This flavor change is a quantum mechanical phenomenon observed in vacuum.  In the presence of matter along the neutrino's path, this probability is altered since neutrinos interact with the matter and scatter. While all flavors scatter via neutral current interactions, this effect is larger for electron neutrinos because they can also partake in charged current interactions with eletrons. These standard matter interactions affect the survival probability of the electron neutrino to a greater extent than that of muon and tau neutrinos, since muon and tau particle densities in normal matter are miniscule compared to electrons. There may be, however, non-standard matter interactions that influence survival probabilities of muon neutrinos.

\subsection{Non-standard Interactions}
If neutrinos participate in non-standard interactions, the flavor Hamiltonian will receive contributions from this effect similar to the standard matter interactions that give rise to the Mikheyev-Smirnov-Wolfenstein (MSW) effect \cite{Kopp:2010qt, Mann:2010jz}. In general, NSI can be flavor changing or flavor conserving, with amplitudes proportional to the standard MSW matter effect $\epsilon_{\mu\mu}V$, $\epsilon_{\tau\tau}V$, and $\epsilon_{\mu\tau}V$, with $V=\sqrt{2}G_{F}N_{e}$. 
\vspace{0.3cm}
\begin{equation}
H_{\text{matter}}=V\left(\begin{array}{ccc}
1 + \epsilon_{ee}0 & \epsilon_{e\mu} & \epsilon_{e\tau}\\
\epsilon_{e\mu}^{*} & \epsilon_{\mu\mu} & \epsilon_{\mu\tau}\\
\epsilon_{e\tau}^{*} & \epsilon_{\mu\tau}^{*} & \epsilon_{\tau\tau}
\end{array}\right).
\end{equation}
\vspace{0.3cm}

MINOS is most sensitive to muon neutrino disappearance and the vacuum oscillation measurements assume a two-flavor model with transitions into tau flavor neutrinos. The limits on $\epsilon_{\mu\mu}$ are already stringent and the difference between $\overline{\nu}_{\mu}$ and $\nu_{\mu}$ oscillations is more sensitive to $\epsilon_{\mu\tau}$ than $\epsilon_{\tau\tau}$ \cite{Mann:2010jz}, therefore the flavor-conserving part is neglected in the following discussion. Only the real part of $\epsilon_{\mu\tau}$ distinguishes neutrinos from antineutrinos; the CP-violating imaginary part is neglected. This gives the following two-flavor NSI Hamiltonian, where $\epsilon_{\mu\tau}$ is the real-valued, flavor changing neutral current contribution to non-standard interactions: 

\vspace{0.3cm}
\begin{equation}
H_{\text{matter}}=\left(\begin{array}{cc}
0 & \epsilon_{\mu\tau}V\\
\epsilon_{\mu\tau}V & 0\end{array}\right).
\end{equation}
\vspace{0.3cm}

\begin{figure}[h]
\begin{center}
  \includegraphics[width=.6\linewidth]{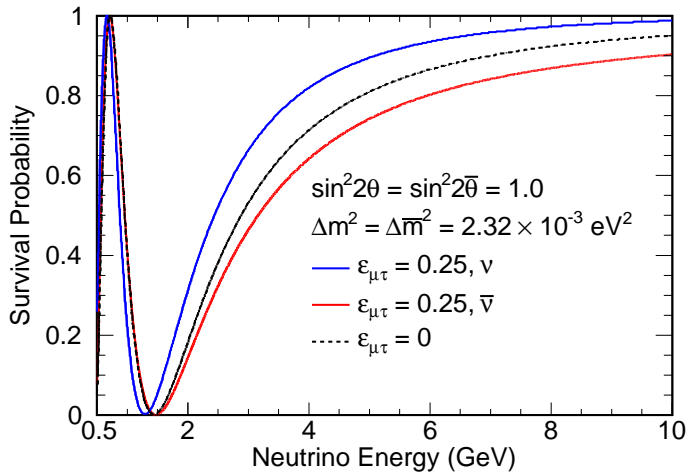}
  \caption{\label{fig:survprob}Survival probability of neutrinos and antineutrinos in the presence of non-standard matter interactions upon traveling 735km from the source. MINOS best fit values for standard oscillations are assumed, \dm\ = $2.32\times10^{-3}$eV$^2$ and \sn\ = 1.0.}
\end{center}
\end{figure}

Using the fact that $V\rightarrow-V$ for antineutrinos we can write the full Hamiltonian, $H=H_{0} + H_{\text{matter}}$ as:

\vspace{0.3cm}
\begin{equation}
H=\left(\begin{array}{cc}
\sin^{2}\theta_{23}\frac{\Delta m^{2}}{2E} & \sin\theta_{23}\cos\theta_{23}\frac{\Delta m^{2}}{2E}\pm\epsilon_{\mu\tau}V\\
\sin\theta_{23}\cos\theta_{23}\frac{\Delta m^{2}}{2E}\pm\epsilon_{\mu\tau}V & \cos^{2}\theta_{23}\frac{\Delta m^{2}}{2E}\end{array}\right)
\end{equation} 
\vspace{0.3cm}

\noindent with which the survival probability becomes, in units of GeV for neutrino energy, km for baseline and eV$^2$ for mass splitting:

\begin{equation}
P = \cos^{2}F_{1} + \frac{\cos^{2}(2\theta)\sin^{2}F_{1}}{F_{2}}
\end{equation}

\noindent where

\begin{eqnarray}
F_{1}  &=& \sqrt{ 
\left(1.27\frac{\Delta m^{2}L}{E}\right)^{2} 
\pm2\sin(2\theta)\left(1.27\frac{\Delta m^{2}L}{E}\right)\epsilon VL
+ (\epsilon VL)^{2}
}\\
F_{2}  &=&  1 \pm \frac{\sin(2\theta)\epsilon VL}{\left(1.27\frac{\Delta m^{2}L}{E}\right)}
+ \left[ \frac{\epsilon VL}{\left(1.27\frac{\Delta m^{2}L}{E}\right)}  \right]^{2} 
\end{eqnarray}
Figure \ref{fig:survprob} shows a comparison of the vacuum survival probability to the probability with non-standard matter interactions at maximal mixing.

%%%%%%%%%%%%%%%%%%%%%%%%%%%%%%%%%%%%%%%%%%%%%%%%%%%%%
\section{The MINOS Experiment}

MINOS is a long baseline neutrino experiment which searches for neutrino oscillations primarily in the atmospheric sector by sending a beam of predominanly muon neutrinos 735 km from Fermilab to the Soudan mine in Minnesota. Two functionally identical detectors measure the neutrino energy spectrum before and after oscillations occur. This allows for a meeasurement of neutrino oscillations by looking for a disappearance between the Near Detector, 1km downstream of the target at Fermilab and the Far Detector, 735 km downstream at the Soudan Mine.

The NuMI (Neutrinos at the Main Injector) high intensity neutrino beam is generated at Fermilab by extracting 120 GeV protons from the Main Injector and striking a graphite target. This produces pions and kaons which are focused into a 675 m long decay pipe by two magnetic focusing horns. Mesons travel down the decay pipe, filled with Helium at 0.9 atm.  Undecayed hadrons and muons coming from the decay of mesons are absorbed and monitored. The resulting beam is made purely of neutrinos and antineutrinos. 

MINOS measures neutrino and antineutrino oscillations directly by utilizing NuMI's ability produce a $\nu_\mu$- or $\overline{\nu}_{\mu}$-enhanced beam.  In neutrino-mode the magnetic field in the horns is adjusted so that positively charged mesons are focused into the decay pipe which decay into neutrinos. In antineutrino-mode negative mesons are selected which enhances the antineutrino composition of the beam, as well as resulting in a pronounced peak at the energy range where oscillations are expected at around 3GeV.

%%%%%%%%%%%%%%%%%%%%%%%%%%%%%%%%%%%%%%%%%%%%%%%%%%%%%
\section{The NSI analysis}
The analysis is performed in several steps: Selection of charged current neutrino and antineutrino events, measurement of the Near Detector and Far Detector energy spectra, calculation of the predicted Far Detector spectrum from the near, and a combined fit to neutrino and antineutrino datasets to measure common oscillation parameters and the NSI parameter, $\epsilon$.

\subsection{Event Selection}
A preselection is applied to all data samples to ensure data quality. The same set of data quality criteria are required for both neutrino and antineutrino samples. We select on good beam type, good horn current and good detector coil current as the running conditions require. We require that events have at least one track and have track vertices in the fiducial volume.

\begin{figure}[h]
\begin{center}
\includegraphics[width=0.5\linewidth]{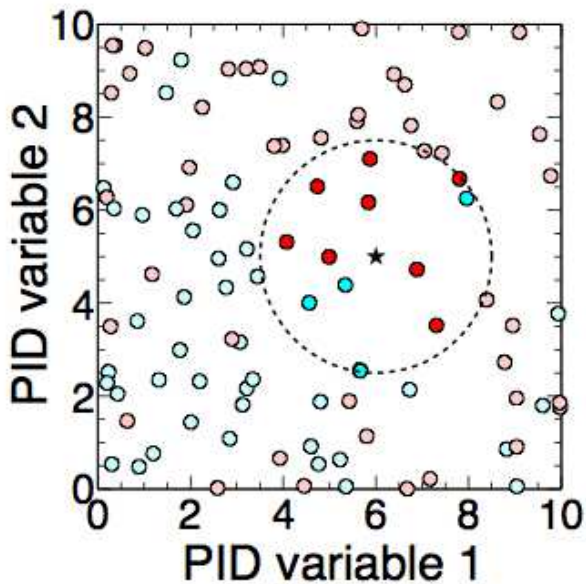}
\caption{\label{fig:kNN}A cartoon demonstrating the use of the $k$NN particle identification in two-dimensional $k$NN space. 12 nearest neighbors lie in the dashed circle, red representing signal and blue representing background. This gives the event a $k$NN PID value of $\frac{2}{3}$}
\end{center}
\end{figure}

Once data quality is satisfied, there are two major sources of background: neutral current events and events of the opposite charge sign. To separate charged current events from neutral currents, we employ a $k$-Nearest Neighbors ($k$NN) algorithm. It is a multivariate method in which an N-dimensional space of track variables is constructed. A query event is identified as signal-like or background-like by its $k$ neighbors' characteristics. The distance in $k$NN space is determined by 

\begin{equation}
D=\left( \sum_{i=1}^{d}\left| X^{MC}_{i}-X^{Q}_{i} \right|^{2} \right)^{\frac{1}{2}}
\end{equation}

\noindent where $MC$ denotes the Monte Carlo events of known type (charged or neutral current) and $Q$ is the query event. The particle identification parameter ($k$NN PID) is then given by

\begin{equation}
k\text{NN}_{\text{ID}}=\frac{k_{S}}{k_{S}+k_{B}}
\end{equation}

\noindent where $S$ and $B$ denote signal and background respectively. In the cartoon in figure \ref{fig:kNN}, red dots denote signal and blue dots background. In this case $k=12$, $k_{S}=8$, $k_{B}=4$ which gives the query event marked by the star a $k$NN$_{\text{ID}}$ of $\frac{2}{3}$.

\begin{figure}[h]
  \begin{center}
    \includegraphics[width=.6\linewidth]{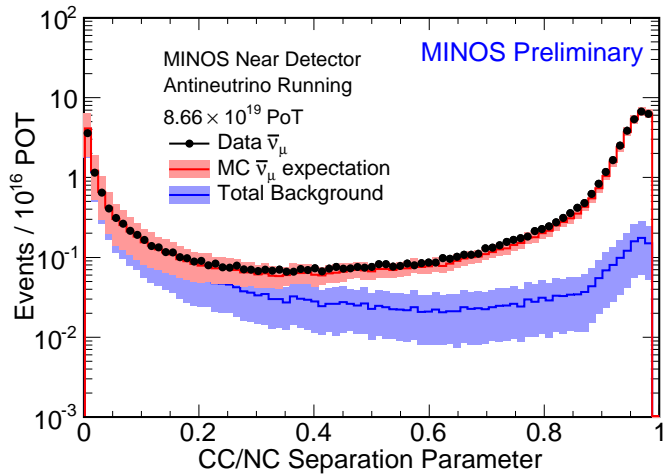}
    \caption{\label{fig:rhc_roID}The $k$NN PID distribution of antineutrino events in the Near Detector. Events above 0.3 are kept.}
  \end{center}
\end{figure}

\begin{figure}[h]
  \begin{center}
    \includegraphics[width=.45\linewidth]{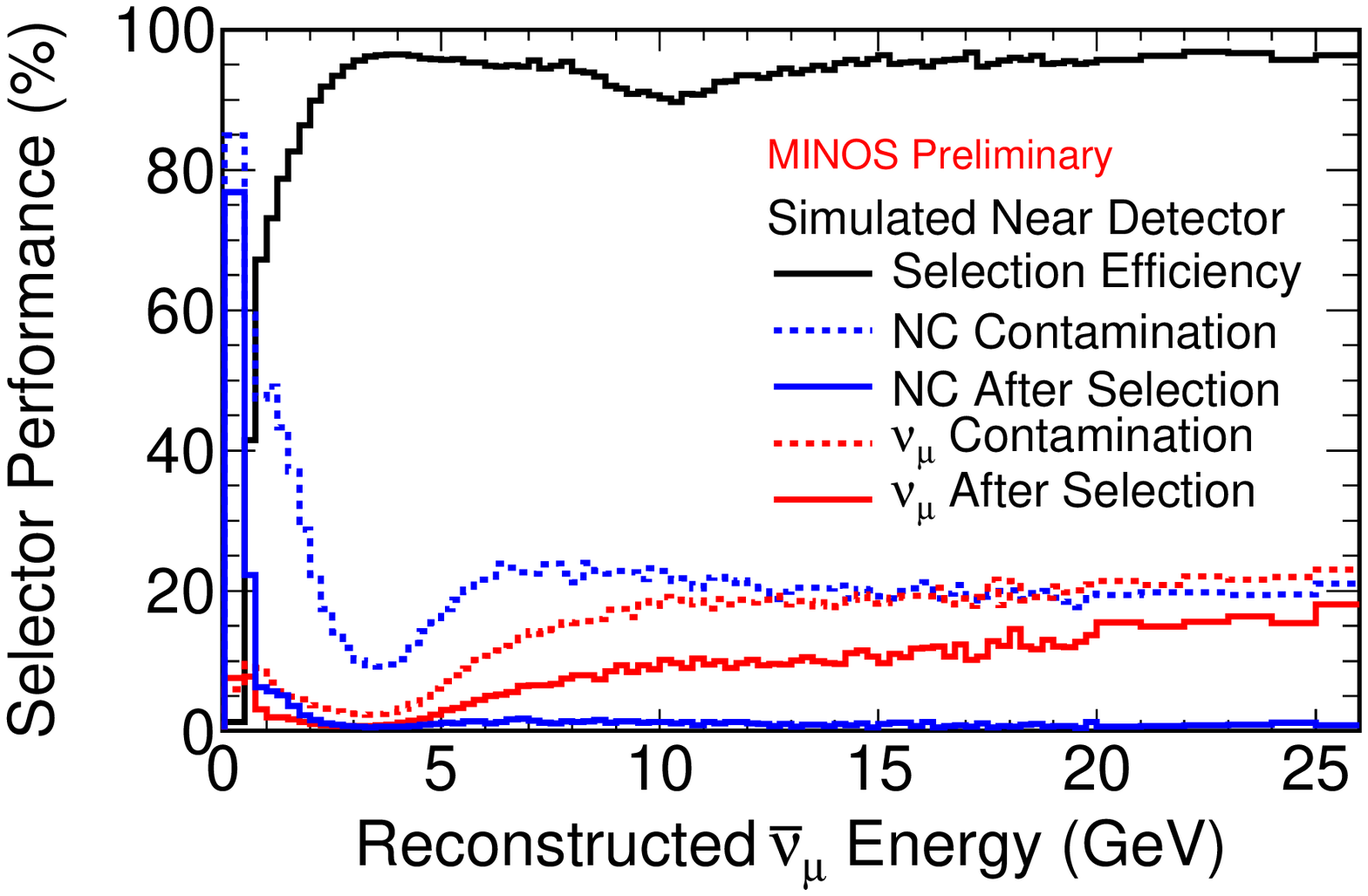}
    \includegraphics[width=.45\linewidth]{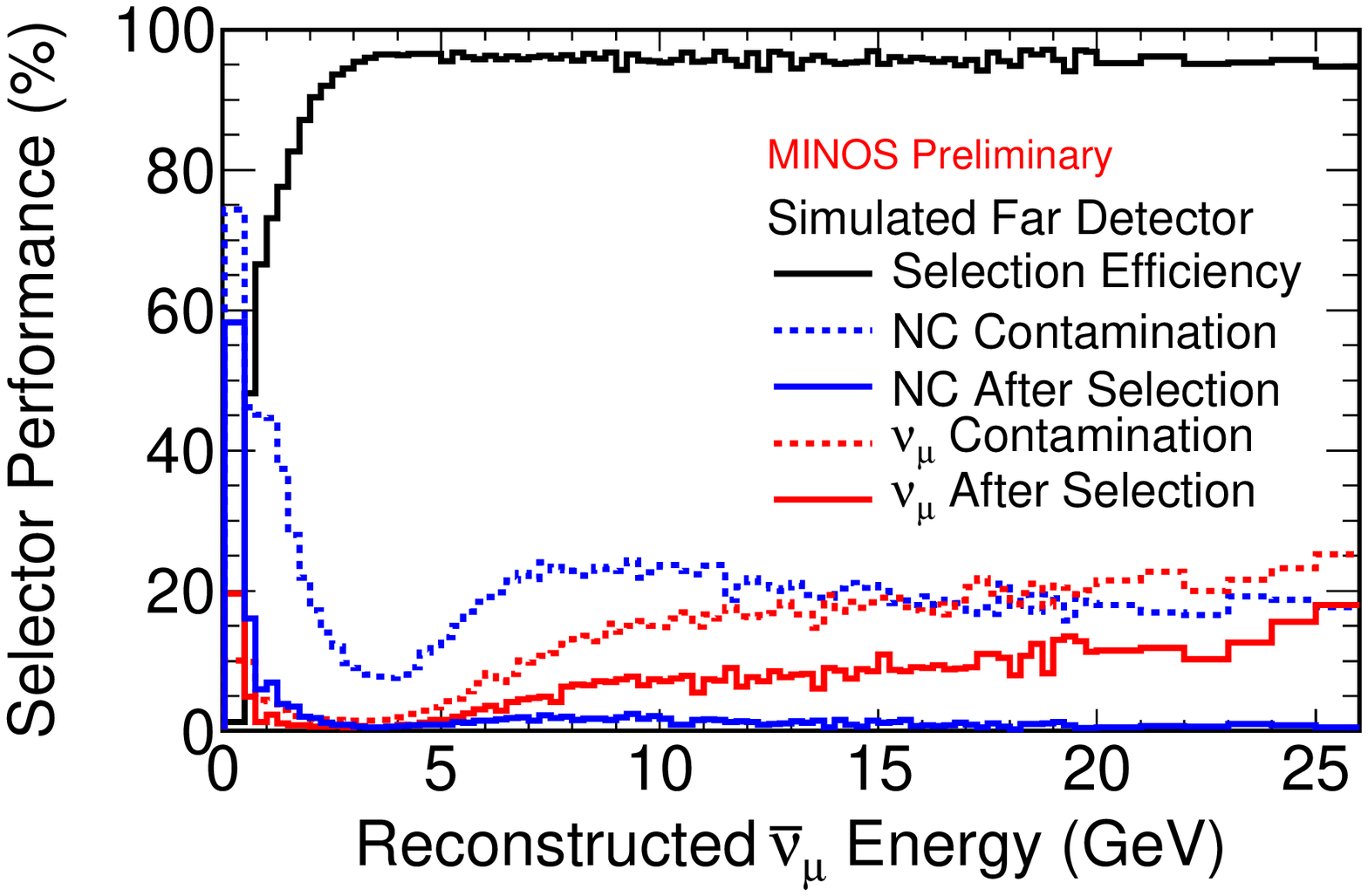}
    \caption{\label{fig:rhcEff}Performance of the antineutrino selection ($kNN_{ID} > 0.3$, $qp > 0$).  The dashed lines show the contamination before selection and the solid show efficiency and contamination after selection.}
  \end{center}
\end{figure}

To select antineutrinos, we keep events with $k$NN $>0.3$. The $k$NN PID distribution of antineutrinos (before selection) is shown in figure \ref{fig:rhc_roID}. We further require that events are reconstructed with positive charge. The efficiency and purity of this selector are shown in figure \ref{fig:rhcEff}. 

\begin{figure}[h]
  \begin{center}
    \includegraphics[width=.45\linewidth]{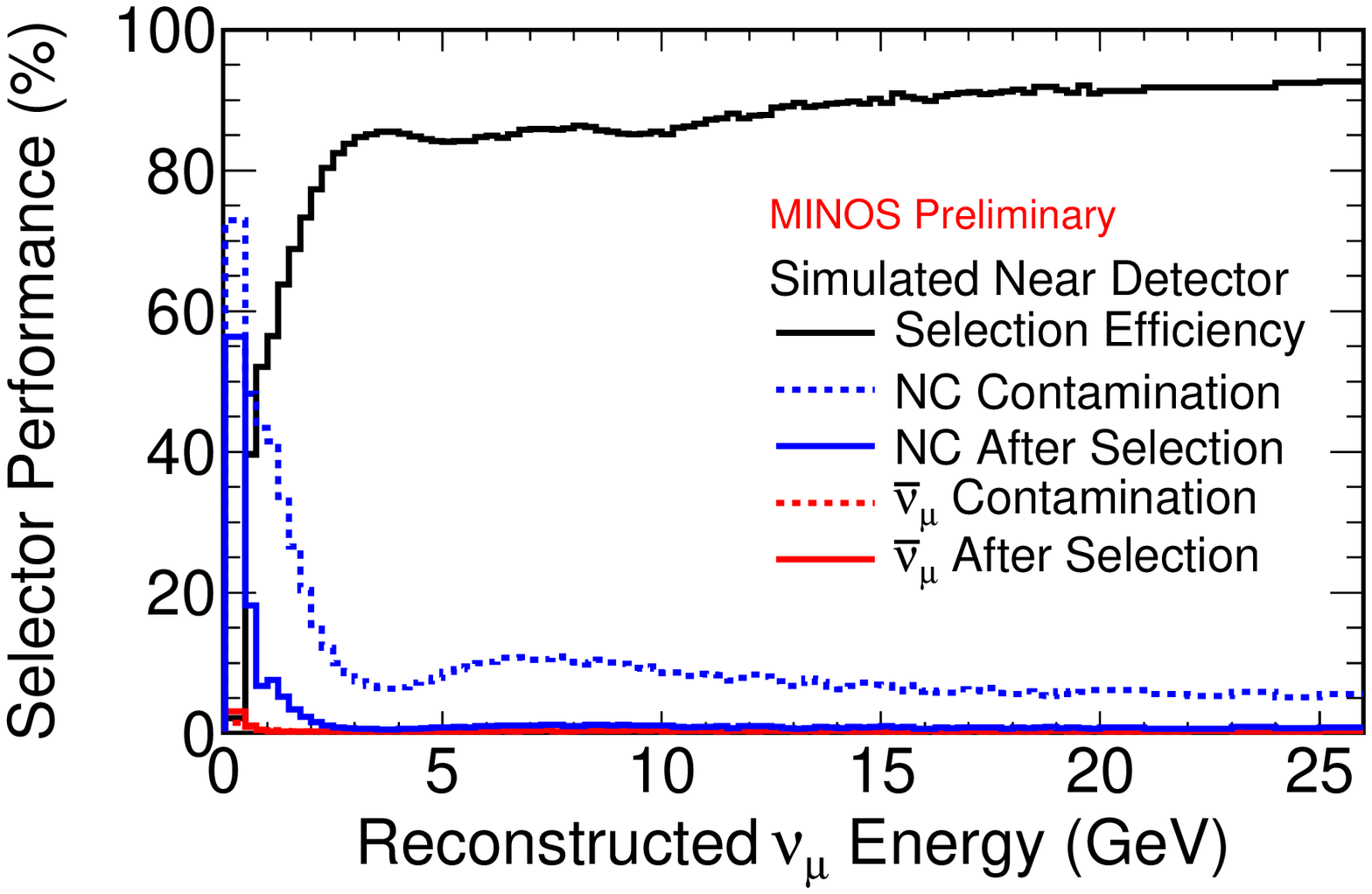}
    \includegraphics[width=.45\linewidth]{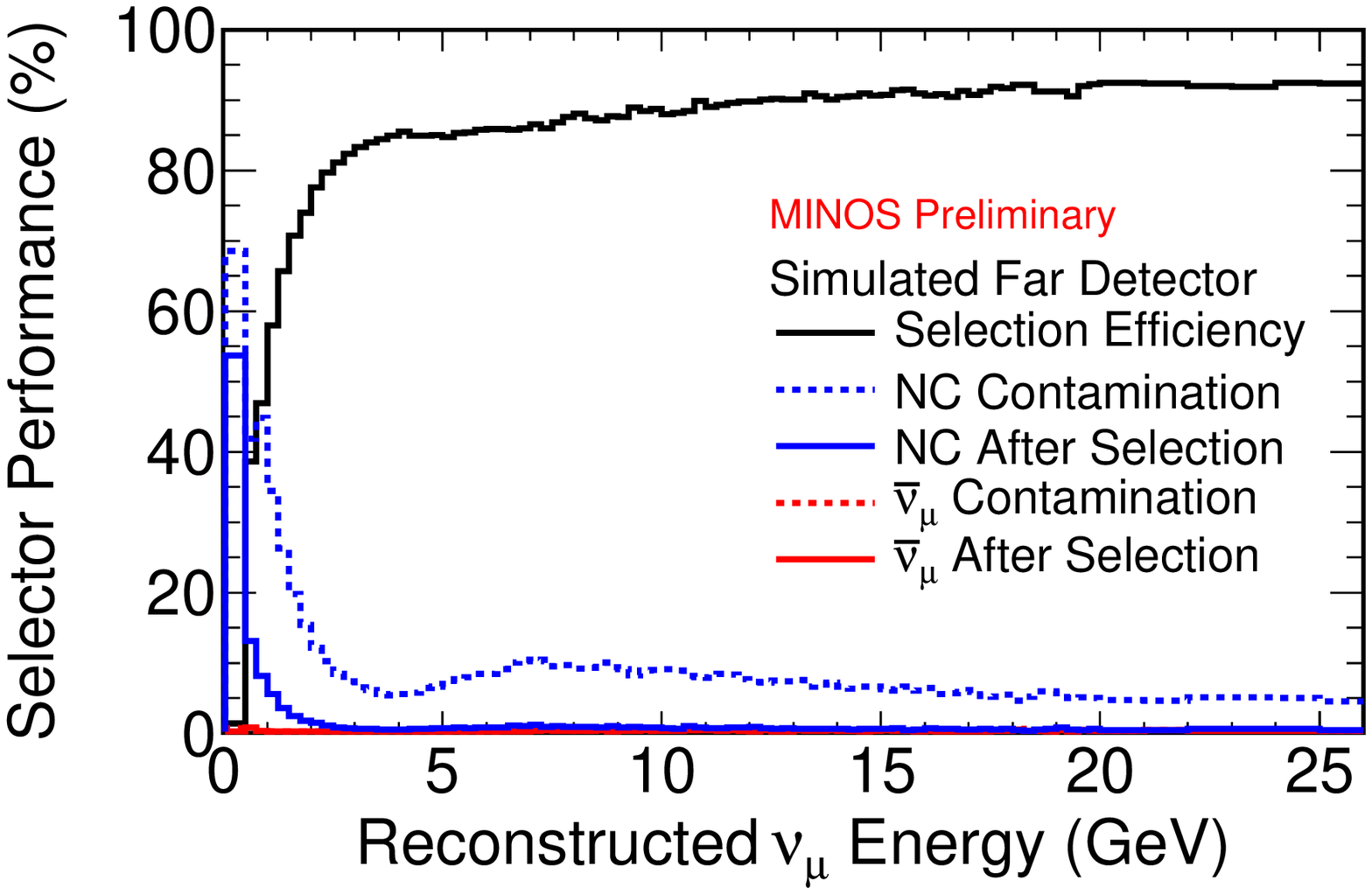}
    \caption{\label{fig:fhcEff}Performance of the neutrino selection.  The dashed lines show the contamination before selection and the solid show efficiency and contamination after selection.}
  \end{center}
\end{figure}

The neutrino selection also uses the $k$NN PID to separate charged current and neutral current events. An additional $k$NN selector which improves low energy selection efficiency is also applied to select neutrino events. The overall neutrino CC/NC separator is a logical \textbf{OR} of these two discriminators. Events are required to have negative reconstructed charge.  The efficiency and purity of the neutrino selection are shown in figure \ref{fig:fhcEff}.

\subsection{Extrapolation}
\begin{figure}[h]
\begin{center}
\includegraphics[width=0.4\linewidth]{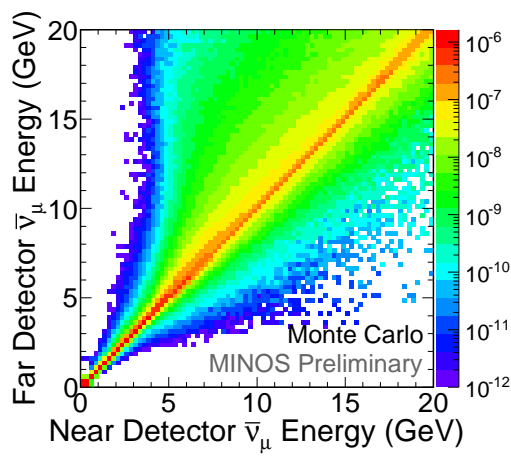}
\caption{\label{fig:beam_matrix} The beam matrix uses Monte Carlo truth information to trace a parent pion to its daughter neutrinos in the Near and Far Detectors. The matrix is then applied to the observed ND data spectrum to obtain the FD predicted spectrum.}
\end{center}
\end{figure}

The Near and Far Detector spectra are similar but not identical even in the absence of oscillations. Due to its proximity to the decay pipe, the Near Detector accepts boosted events that decay further downstream in the pipe whereas the Far Detector views it as a point source. To obtain an accurate prediction which takes into account these effects, a beam matrix is used. This matrix, shown in figure \ref{fig:beam_matrix}, is obtained by tracing a parent pion of known energy from Monte Carlo to its daughter neutrinos in the Near and Far Detectors. This matrix then muultiplies the measured Near Detector data spectrum to obtain a Far Detector prediction in a data-driven manner. Neutrinos and antineutrinos are extrapolated separately with their respective beam matrices, the figure shows the antineutrino matrix. 

\subsection{Far Detector Data and Fitting}
\begin{figure}[h]
\begin{center}
\includegraphics[width=.45\linewidth]{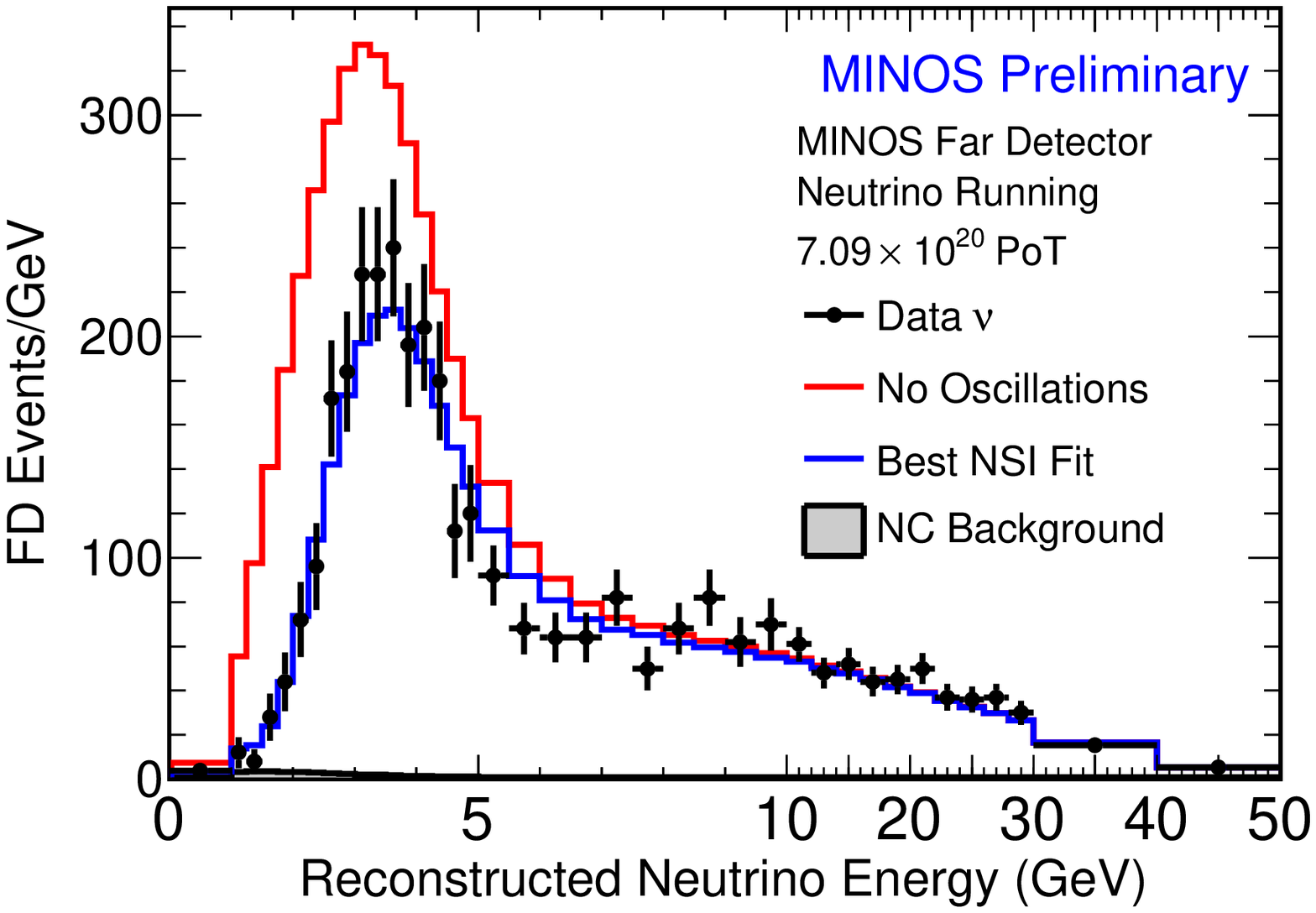}
\includegraphics[width=.45\linewidth]{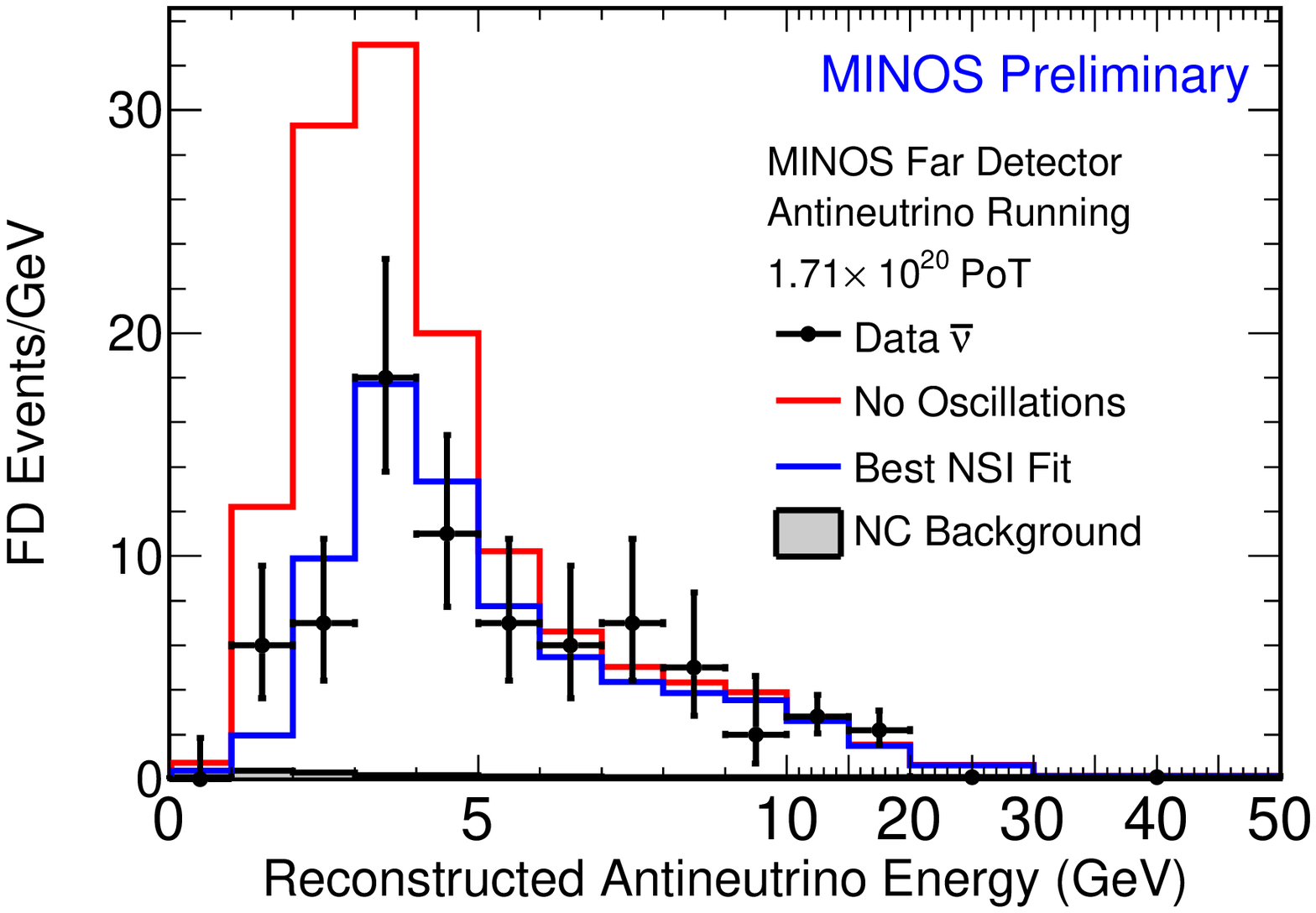}\\
\includegraphics[width=.45\linewidth]{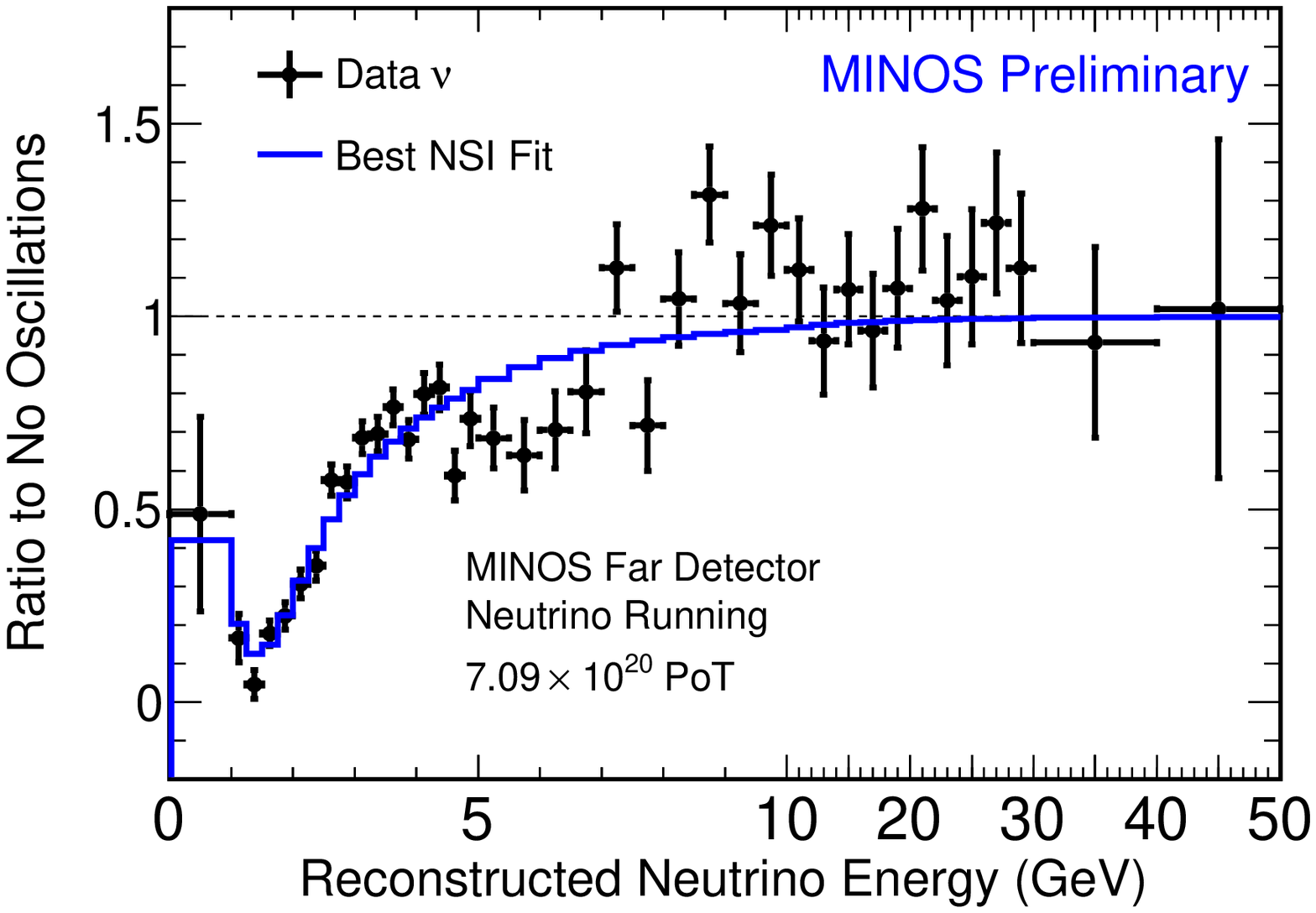}
\includegraphics[width=.45\linewidth]{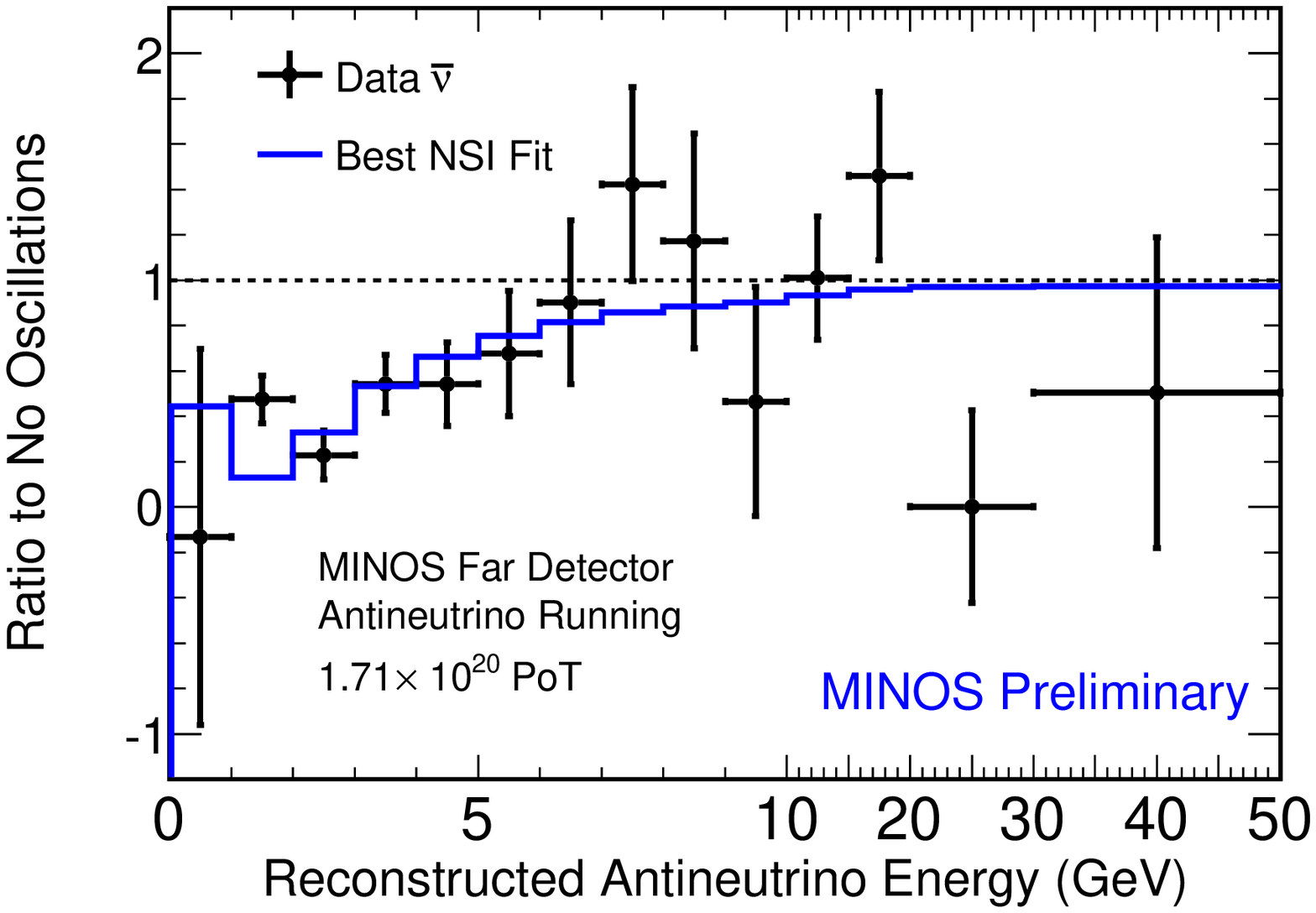}
\caption{\label{fig:spectra}Far Detector reconstructed neutrino energy spectra of neutrino (left) and antineutrino samples. The red line shows the predicted spectrum without oscillations while the blue and green curves are fits to non-standard interactions and vacuum oscilations respectively. The bottom plots show the ratio of oscillated to unoscillated spectra.}
\end{center}
\end{figure}

We expect $2073$ neutrino and $156$ antineutrino events at the Far Detector in the absence of oscillations, and observe $1654$ and $97$ respectively. We fit for non-standard interactions in 100 bins per sample (for a total of 400 bins in three neutrino and one antineutrino run periods). The effect of the four largest systematic uncertainties are included in the fit as penalty terms. Figure \ref{fig:spectra} shows the Far Detector energy spectra of neutrinos and antineutrinos. 

\begin{figure}[h]
\begin{center}
\includegraphics[width=.32\linewidth]{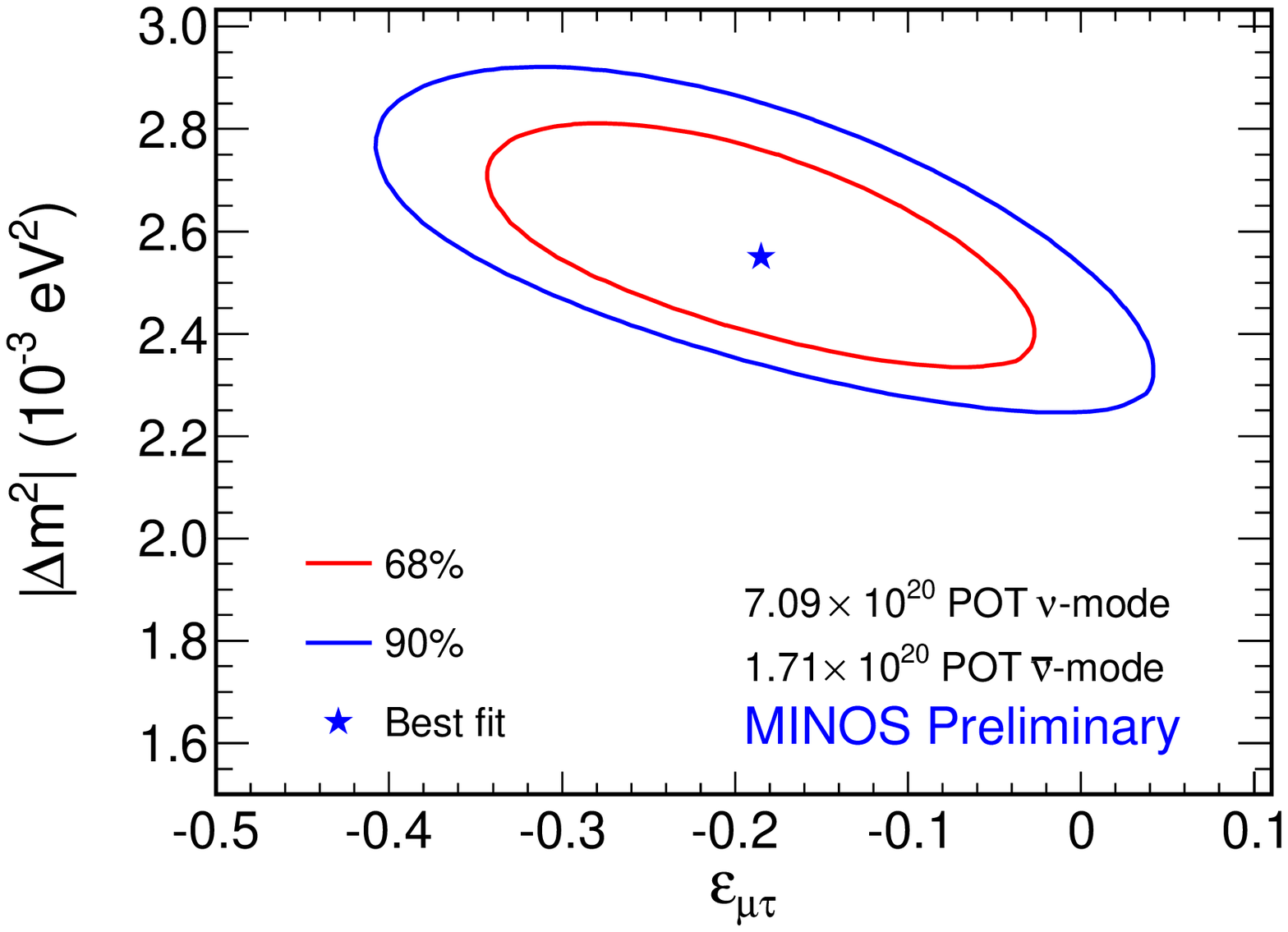}
\includegraphics[width=.32\linewidth]{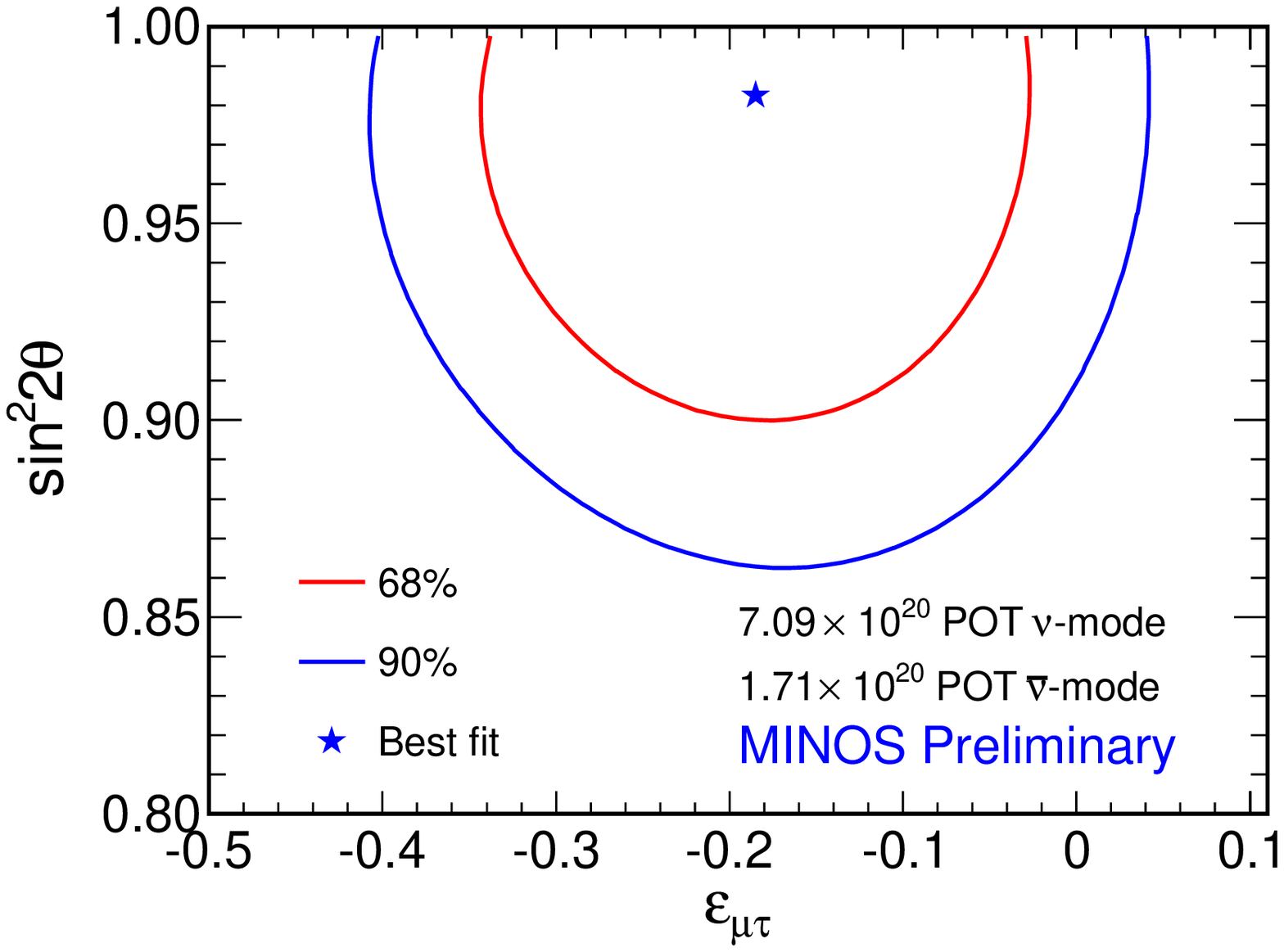}
\includegraphics[width=.32\linewidth]{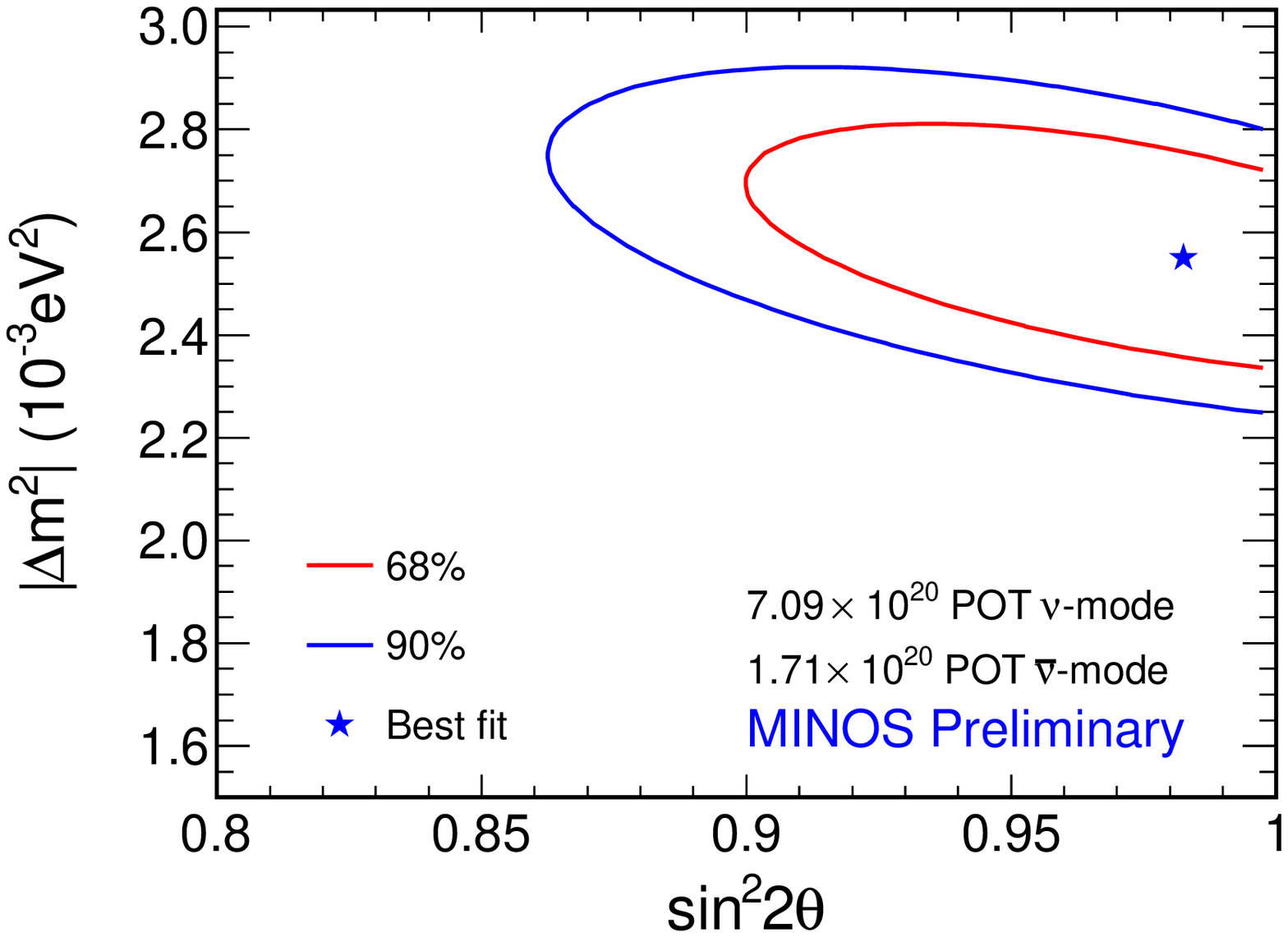}
\caption{\label{fig:contours}Allowed regions for \dm, \sn, and \eps. Contours assume Gaussian statistics and include the effect of systematics.}
\end{center}
\end{figure}

The fit to non-standard interactions yields \dm$=2.56\pm0.15\times 10^{-3}$ eV$^{2}$, \sn$>0.90$, and \eps=$-0.187\pm0.16$. The confidence intervals are shown in figure \ref{fig:contours}.

\begin{acknowledgements}
We thank DPF and the organizers for the opportunity to present these results at its 2011 meeting. These proceedings report work supported by the U.S. DOE; the U.K. STFC; the U.S. NSF; the State and University of Minnesota; the University of Athens, Greece; and Brazil’s FAPESP, CNPq, and CAPES. We are grateful to the Minnesota Department of Natural Resources, the crew of the Soudan Underground Laboratory, and the staff of Fermilab for their contributions to this effort.
\end{acknowledgements}

\newpage
\bibliography{bibliography.bib}
\end{document}